\documentclass[aps,prd,a4paper,twocolumn,amsmath,showpacs,superscriptaddress,nofootinbib,preprintnumbers]{revtex4-1}
\pdfoutput=1
\usepackage{amssymb,amsmath,latexsym,mathrsfs}
\usepackage{graphicx,subfigure}
\usepackage{epsfig}
\usepackage{varioref,xr-hyper}
\usepackage{color}
\usepackage{multirow}
\usepackage{array}
\usepackage{hyperref}
\usepackage{wasysym}
\usepackage{color}
\usepackage{float}
\usepackage{xcolor}
\usepackage[utf8]{inputenc}
\usepackage[T1]{fontenc}

\begin{document}

\preprint{LCTP-18-18,NORDITA-2018-053,TTK-18-24}

\title{Bias due to neutrinos must not uncorrect'd go}

\author{Sunny Vagnozzi}
\email{sunny.vagnozzi@fysik.su.se}
\affiliation{The Oskar Klein Centre for Cosmoparticle Physics, Department of Physics, Stockholm University, SE-106 91 Stockholm, Sweden}
\affiliation{The Nordic Institute for Theoretical Physics (NORDITA), Roslagstullsbacken 23, SE-106 91 Stockholm, Sweden}

\author{Thejs Brinckmann}
\email{brinckmann@physik.rwth-aachen.de}
\affiliation{Institute for Theoretical Particle Physics and Cosmology (TTK), RWTH Aachen University, Otto-Blumenthal Stra{\ss}e, D-52056 Aachen, Germany}

\author{Maria Archidiacono}
\affiliation{Institute for Theoretical Particle Physics and Cosmology (TTK), RWTH Aachen University, Otto-Blumenthal Stra{\ss}e, D-52056 Aachen, Germany}

\author{Katherine Freese}
\affiliation{The Oskar Klein Centre for Cosmoparticle Physics, Department of Physics, Stockholm University, SE-106 91 Stockholm, Sweden}
\affiliation{The Nordic Institute for Theoretical Physics (NORDITA), Roslagstullsbacken 23, SE-106 91 Stockholm, Sweden}
\affiliation{Leinweber Center for Theoretical Physics, Department of Physics, University of Michigan, Ann Arbor, MI 48109, USA}

\author{Martina Gerbino}
\affiliation{The Oskar Klein Centre for Cosmoparticle Physics, Department of Physics, Stockholm University, SE-106 91 Stockholm, Sweden}

\author{Julien Lesgourgues}
\affiliation{Institute for Theoretical Particle Physics and Cosmology (TTK), RWTH Aachen University, Otto-Blumenthal Stra{\ss}e, D-52056 Aachen, Germany}

\author{Tim Sprenger}
\affiliation{Institute for Theoretical Particle Physics and Cosmology (TTK), RWTH Aachen University, Otto-Blumenthal Stra{\ss}e, D-52056 Aachen, Germany}

\date{\today}

\begin{abstract}
It is a well known fact that galaxies are biased tracers of the distribution of matter in the Universe. The galaxy bias is usually factored as a function of redshift and scale, and approximated as being scale-independent on large, linear scales. In cosmologies with massive neutrinos, the galaxy bias defined with respect to the total matter field (cold dark matter, baryons, and non-relativistic neutrinos) also depends on the sum of the neutrino masses $M_{\nu}$, and becomes scale-dependent even on large scales. This effect has been usually neglected given the sensitivity of current surveys. However, it becomes a severe systematic for future surveys aiming to provide the first detection of non-zero $M_{\nu}$. The effect can be corrected for by defining the bias with respect to the density field of cold dark matter and baryons, rather than the total matter field. In this work, we provide a simple prescription for correctly mitigating the neutrino-induced scale-dependent bias effect in a practical way. We clarify a number of subtleties regarding how to properly implement this correction in the presence of redshift-space distortions and non-linear evolution of perturbations. We perform a Markov Chain Monte Carlo analysis on simulated galaxy clustering data that match the expected sensitivity of the \textit{Euclid} survey. We find that the neutrino-induced scale-dependent bias can lead to important shifts in both the inferred mean value of $M_{\nu}$, as well as its uncertainty, and provide an analytical explanation for the magnitude of the shifts. We show how these shifts propagate to the inferred values of other cosmological parameters correlated with $M_{\nu}$, such as the cold dark matter physical density $\Omega_{cdm} h^2$ and the scalar spectral index $n_s$. In conclusion, we find that correctly accounting for the neutrino-induced scale-dependent bias will be of crucial importance for future galaxy clustering analyses. We encourage the cosmology community to correctly account for this effect using the simple prescription we present in our work. The tools necessary to easily correct for the neutrino-induced scale-dependent bias will be made publicly available in an upcoming release of the Boltzmann solver \texttt{CLASS}.
\end{abstract}

\maketitle

\section{Introduction}
\label{sec:introduction}

Neutrinos are one of the most abundant particle species in the Universe, yet remain among the least understood. While the Standard Model of Particle Physics treats neutrinos as fundamental massless particles, neutrino oscillation experiments have shown that at least two out of the three neutrino mass eigenstates are massive. Therefore, massive neutrinos represent the only direct evidence for physics beyond the Standard Model. In fact, neutrino oscillation experiments have measured two mass-squared splittings: the solar splitting $\Delta m_{21}^2 \equiv m_2^2-m_1^2 \simeq 7.6 \times 10^{-5}\,{\rm eV}^2$ and the atmospheric splitting $\vert \Delta m_{31}^2 \vert \equiv \vert m_3^2-m_1^2 \vert \simeq 2.5 \times 10^{-3}\,{\rm eV}^2$~\cite{Gonzalez-Garcia:2015qrr,Capozzi:2016rtj,Esteban:2016qun,Capozzi:2017ipn,deSalas:2017kay}, where $m_1$, $m_2$, $m_3$ denote the masses of the three mass eigenstates. The uncertainty in the sign of the atmospheric splitting leaves two possibilities open for the neutrino mass ordering: a \textit{normal ordering} (NO) where $\Delta m_{31}^2 >0$ and $m_1<m_2<m_3$, and an \textit{inverted ordering} (IO) where $\Delta m_{31}^2<0$ and $m_3<m_1<m_2$.

The existence of a cosmic background of relic neutrinos is a prediction of the standard cosmological model. Cosmic neutrinos are initially in equilibrium with the primordial plasma. They decouple at temperatures around $T\sim\,{\rm MeV}$, while still relativistic, and start to free-stream. When the temperature of the Universe drops below the temperature equivalent to the mass of a neutrino species, neutrinos turn non-relativistic and start contributing to the matter content of the Universe. This unique behaviour of massive neutrinos during the expansion history of the Universe leaves several peculiar signatures in cosmological observables (see e.g.~\cite{Lesgourgues:2006nd,Wong:2011ip,Lesgourgues:1519137,Lattanzi:2017ubx} for comprehensive reviews on neutrino cosmology). By means of these signatures, cosmology is sensitive to the sum of the neutrino masses $M_{\nu}$:
\begin{eqnarray}
M_{\nu} \equiv m_1+m_2+m_3 \, .
\end{eqnarray}

Cosmological data currently provide the tightest upper limits on $M_{\nu}$, suggesting $M_{\nu} \lesssim 0.15\,{\rm eV}$ at $95\%$ confidence level (C.L.)~\cite{Ade:2015xua,Palanque-Delabrouille:2015pga,DiValentino:2015wba,Cuesta:2015iho,Huang:2015wrx,Moresco:2016nqq,Giusarma:2016phn,Alam:2016hwk,Vagnozzi:2017ovm,Capozzi:2017ipn,
Couchot:2017pvz,Caldwell:2017mqu,Doux:2017tsv,Wang:2017htc,Chen:2017ayg,Upadhye:2017hdl,Salvati:2017rsn,Nunes:2017xon,Zennaro:2017qnp,Wang:2018lun,
Choudhury:2018byy,Choudhury:2018adz,Aghanim:2018eyx}, with potentially interesting implications concerning the determination of the mass ordering~\cite{Hannestad:2016fog,Xu:2016ddc,Gerbino:2016ehw,Vagnozzi:2017ovm,Simpson:2017qvj,Schwetz:2017fey,Hannestad:2017ypp,
Long:2017dru,Gariazzo:2018pei,Heavens:2018adv,Handley:2018gel,deSalas:2018bym}. These bounds are somewhat model-dependent and rely on assuming an underlying background standard $\Lambda$CDM cosmological model. The upper limits on $M_{\nu}$ typically degrade when considering extensions of this baseline model (e.g. when the dark energy equation of state is allowed to vary), see e.g.~\cite{Hannestad:2005gj,Joudaki:2012fx,Vagnozzi:2017ovm,Yang:2017amu,Lorenz:2017fgo,Wang:2017htc,Chen:2017ayg,Sutherland:2018ghu,
Choudhury:2018byy,Choudhury:2018adz,Sahlen:2018cku} for recent investigations, although there are exceptions to this statement: for instance, allowing for a time-varying dark energy component, but forcing it to lie in the \textit{non-phantom} region (i.e. $w(z) \geq -1$), actually results in tighter upper limits on $M_{\nu}$~\cite{Vagnozzi:2018jhn}. At any rate, moving from these upper limits to a robust detection of $M_{\nu}$ is a key goal of upcoming cosmological surveys, as it would provide a unique window into physics beyond the Standard Model. It is expected that a combination of near-future cosmological observations should indeed lead to the first detection of non-zero $M_{\nu}$~\cite{Lesgourgues:2004ps,Carbone:2010ik,Hamann:2012fe,Audren:2012vy,Allison:2015qca,Hlozek:2016lzm,Abazajian:2016yjj,Archidiacono:2016lnv,
DiValentino:2016foa,Lattanzi:2017ubx,Sprenger:2018tdb,Mishra-Sharma:2018ykh,Brinckmann:2018owf,Ade:2018sbj}.

The free-streaming nature of neutrinos underlies one of their most peculiar cosmological signatures. We can define a free-streaming scale as the scale covered by a free-streaming neutrino during one Hubble time. Below the free-streaming scale, neutrinos are unable to cluster due to their large thermal velocity. Moreover, massive neutrinos slow down the growth of matter perturbations. These effects result in a small-scale suppression of the matter power spectrum $P_m(k,z)$. The clustering of large-scale structure is therefore a potentially sensitive probe of this effect, and hence of $M_{\nu}$.

Considerable effort is being devoted towards future large-scale structure surveys, such as galaxy redshift surveys, able to access smaller and increasingly non-linear scales. At the same time, achieving a reliable detection of $M_{\nu}$ requires an exquisite control of systematics, including both instrumental and modeling systematics. A potential particularly delicate modeling systematic stems from the fact that we only have access to \textit{biased} tracers of the matter power spectrum, such as the clustering of galaxies. The underlying matter power spectrum is related to the measured power spectrum of a biased tracer $P_t(k,z)$ as follows:
\begin{eqnarray}
P_t(k,z) = b_m^2(k,z)P_m(k,z) \, ,
\label{eq:powerspectrum}
\end{eqnarray}
where the bias $b_m$ quantifies the statistical relation between the clustering of matter and of its luminous tracers (see e.g.~\cite{Desjacques:2016bnm} for a recent review). In this work we will be concerned with galaxies as biased tracers of the underlying matter field. 

The bias $b_m$ as appearing in Eq.~(\ref{eq:powerspectrum}) is defined with respect to the \textit{total} matter field, comprising cold dark matter (CDM), baryons, and massive neutrinos. When considering galaxies as tracers, already at this point we can appreciate an important subtlety. In fact, Eq.~(\ref{eq:powerspectrum}) implicitly assumes that galaxies trace the \textit{total} matter field. However, the typical scales probed by galaxy clustering are below the neutrino free-streaming scale, where neutrinos do not cluster. Hence, one should actually expect galaxies to trace the CDM+baryons field instead of the total matter field. This expectation has been proved to be valid by several dedicated simulations~\cite{Villaescusa-Navarro:2013pva,Castorina:2013wga,Costanzi:2013bha,Castorina:2015bma} as well as theoretical studies~\cite{Biagetti:2014pha}, and its consequences for parameter inference are the topic of investigation of this work. Our work is not the first one to investigate the impact of this aspect on parameter inference. As far as we are aware, such a study was first carried out in~\cite{Raccanelli:2017kht}, which found that taking this effect into account when analysing future data will be crucial. In our work we confirm this finding, although there are a number of differences between our analysis and that of~\cite{Raccanelli:2017kht}, which we shall comment more on later~\footnote{Another interesting avenue pursued recently has been the study of void bias, with voids traced by the CDM field, in massive neutrinos cosmologies~\cite{Kreisch:2018var}.}.

In cosmologies \textit{without} massive neutrinos, the behaviour of the bias as a function of scale (wavenumber $k$) is well understood, albeit increasingly harder to model at smaller scales. On large, linear scales ($k\lesssim0.15\,h\,{\rm Mpc}^{-1}$ at $z=0$), the bias can be modeled to very good approximation as being constant~\cite{Desjacques:2016bnm}. On smaller, non-linear scales, complexities inherent to the process of galaxy formation and evolution make the bias intrinsically scale-dependent~\cite{Desjacques:2016bnm,Seljak:2000jg}. On mildly non-linear scales, accurate phenomenological parametrizations of the scale-dependent bias exist (see e.g.~\cite{Amendola:2015pha} for constraints on scale-dependent galaxy bias models from current data, and~\cite{Giusarma:2018jei} for a study on the scale-dependent galaxy bias from CMB lensing-galaxy cross-correlations, with applications to constraints on $M_{\nu}$).

In cosmologies \textit{with} massive neutrinos, the situation is more complex. In this case, the bias defined with respect to the \textit{total} matter field is scale-dependent even on \textit{large scales}, with the form of the scale-dependence dictated by the value of $M_{\nu}$: $b_m(k,z)=b_m(k,z,M_{\nu})$. On the other hand, one can define the bias with respect to the CDM+baryons power spectrum $P_{cb}(k)$:
\begin{eqnarray}
P_t(k,z) = b_{cb}^2(k,z)P_{cb}(k,z) \, .
\label{eq:pt_cb}
\end{eqnarray}
In this case, the bias $b_{cb}$ becomes scale-independent on large scales, and \textit{universal}~\cite{Villaescusa-Navarro:2013pva,Castorina:2013wga,Costanzi:2013bha,Castorina:2015bma}. That is, its scale-dependence is no longer determined by $M_{\nu}$, but only by properties inherent to galaxy formation and evolution, and appears only on small scales~\footnote{In reality, even defining the bias with respect to the CDM+baryons field still leaves a residual scale-dependence in $b_{cb}$ on large scales~\cite{LoVerde:2014pxa,LoVerde:2014rxa,LoVerde:2016ahu,Chiang:2017vuk,Munoz:2018ajr}. Essentially, the reason is that the scale-dependent growth induced by massive neutrinos affects the critical collapse overdensity for halos at different scales. Heuristically, we can think of this effect as being due to the impact of massive neutrinos on halo formation, whereas the effect we will be concerned with in this paper is related to whether or not halos can form in first place. For simplicity, since the residual scale-dependence due to the effect of neutrinos on halo collapse is small, we choose to ignore it here. This effect is nonetheless worth exploring in much more detail, and we plan to return to this issue in a future project.}. Hereafter, all quantities (power spectra, biases, and so on) are implicitly considered to be redshift-dependent. Hence, for simplicity, we will drop all $z$-dependences in our equations from now on.

The common approach when deriving constraints on $M_{\nu}$ from clustering measurements is to define the bias with respect to the total matter power spectrum, as in Eq.~(\ref{eq:powerspectrum}). The bias is still approximated as scale-independent, at least on large scales. While formally incorrect, this choice does not affect analyses given the current sensitivity of cosmological data. However, this could be no longer true with upcoming high-precision cosmological data aiming to measure $M_{\nu}$. Therefore, it is timely to explore the impact of neglecting the neutrino-induced scale-dependent bias (NISDB) in light of future clustering measurements.

In this work, we revisit the issue of the neutrino-induced scale-dependent bias in light of future data from the \textit{Euclid} satellite~\cite{Amendola:2012ys,Amendola:2016saw}. As we noted earlier, a previous similar study was performed in~\cite{Raccanelli:2017kht}, where a Fisher matrix approach was adopted. Our work differs from~\cite{Raccanelli:2017kht} on at least three main aspects. Firstly, we use a Markov Chain Monte Carlo (MCMC) approach, which is better suited to capture important parameter correlations, such as those involved in the measurement of the sum of the neutrino masses. The MCMC approach enables us to definitively show that \textit{correcting for the NISDB will be important, in order not to bias the determination of cosmological parameters, in particular the sum of the neutrino masses}. Second, we clarify a number of subtle issues concerning improper ways of correcting for the NISDB effect. Finally, our MCMC forecasts includes an extensive modeling of systematic effects for the upcoming \textit{Euclid} survey.

This paper is organized as follows. In Sec.~\ref{sec:correction}, we show how to correct for the NISDB in practice, and clarify a number of subtle issues concerning this correction. In Sec.~\ref{sec:datasets}, we describe the simulated datasets we consider and how we model them. In Sec.~\ref{sec:results}, we present our findings. They indicate that correcting for the NISDB will be crucial for future cosmological surveys and are summarised in Tab.~\ref{tab:bounds} for the reader's convenience. Finally, in Sec.~\ref{sec:conclusions} we provide concluding remarks. We suggest that the busy reader who is already familiar with the topic in question might skip to Sec.~\ref{subsec:recap} and Sec.~\ref{sec:conclusions}, and take a look at Fig.~\ref{fig:mnu} and Fig.~\ref{fig:tri}.

\section{Bias correction}
\label{sec:correction}

In this section, we describe in detail our prescription for correcting for the NISDB effect. We begin in Sec.~\ref{subsec:norsd} by considering a simple example to get a feeling for how the correction \textit{could} be implemented. In Sec.~\ref{subsec:rsd} we discuss a more general and correct approach in the presence of redshift-space distortions (RSD). We then comment on the impact of non-linear evolution and how it modifies our correction in Sec.~\ref{subsec:non-linear}. Finally, we give a schematic summary of our final prescription for correcting for the NISDB effect, taking into account RSD and non-linear evolution, in Sec.~\ref{subsec:recap}. The busy reader can therefore skip to Sec.~\ref{subsec:recap}.

\subsection{Bias correction in the absence of RSD}
\label{subsec:norsd}

In the absence of RSD, we can compare the two expressions for the power spectrum of a given tracer $P_t(k)$, Eqs.~(\ref{eq:powerspectrum},~\ref{eq:pt_cb}), and obtain the following relation between the ``meaningful'' bias defined with respect to the CDM+baryons field $b_{cb}$ and the ``effective'' one defined with respect to the total matter field $b_m$:
\begin{eqnarray}
b_m(k,M_{\nu})&=&b_{cb}(k)\sqrt{\frac{P_{cb}(k,M_{\nu})}{P_m(k,M_{\nu})}}\, .
\label{eq:bm1}
\end{eqnarray}
At the level of linear perturbation theory, this can be expressed as:
\begin{eqnarray}
b_m(k,M_{\nu})&=&b_{cb}(k)\frac{T_{cb}(k,M_{\nu})}{T_m(k,M_{\nu})}\, ,
\label{eq:bm}
\end{eqnarray}
where $T_{cb}(k,M_{\nu})$ and $T_m(k,M_{\nu})$ are the linear transfer functions of the CDM+baryons and total matter components, respectively. In particular, the CDM+baryons transfer function is given by:
\begin{eqnarray}
T_{cb}(k,M_{\nu}) = \frac{\Omega_{cdm} T_{cdm}(k,M_{\nu})+\Omega_b T_b(k,M_{\nu})}{\Omega_{cdm} + \Omega_b} \, ,
\end{eqnarray}
where $\Omega_{cdm}$ [$\Omega_b$] and $T_{cdm}(k,M_{\nu})$ [$T_b(k,M_{\nu})$] denote the CDM [baryon] density parameter and transfer function, respectively. Notice that the definition of transfer function we adopt automatically includes the growth function $D(k,z)$. In Eq.~(\ref{eq:bm}), we have explicitly written all $k$ and $M_{\nu}$ dependences to emphasize the important fact that $b_{cb}$ is \textit{universal} (its scale-dependence is not dependent on the value of $M_{\nu}$), while $b_m$ is not. The dependence of $b_m$ on $M_{\nu}$ arises from the dependence of the transfer functions on the neutrino mass~\footnote{We remind the reader that the scale-dependence of $b_{cb}$ we explicitly introduced is relevant only on small non-linear scales~\cite{Desjacques:2016bnm}, or equivalently at large $k$, which we will not be concerned with in this work.}.

Equation (\ref{eq:bm}) allows us to express the galaxy power spectrum $P_{t}$ in terms of the total matter power spectrum $P_m$ through a rescaling by a factor $(T_{cb}/T_m)^2$, since one can insert Eq.~(\ref{eq:bm}) into Eq.~(\ref{eq:powerspectrum}) to get:
\begin{eqnarray}
P_t(k,M_{\nu})=b_{cb}^2(k) \left ( \frac{T_{cb}(k,M_{\nu})}{T_m(k,M_{\nu})} \right )^2P_m(k,M_{\nu}) \, .
\label{eq:pt_cb2}
\end{eqnarray}
Notice that all the terms appearing on the right-hand side of Eq.~(\ref{eq:pt_cb2}) are known: the linear total matter power spectrum $P_m$ and  transfer functions $T_{cb}$ and $T_m$ can be computed by using Einstein-Boltzmann codes, and we can model the universal bias $b_{cb}$ accurately up to mildly non-linear scales.

\subsection{The impact of RSD}
\label{subsec:rsd}

Redshift-space distortions (RSD) arise from the fact that galaxies are observed not in real space, but in redshift space. Thus, the redshift-space power spectrum we observe needs to be corrected for the effect of peculiar velocities. Kaiser showed in a seminal paper~\cite{Kaiser:1987qv} that on large scales (small $k$) Eq.~(\ref{eq:powerspectrum}) when not including massive neutrinos is modified due to the effect of RSD to:
\begin{eqnarray}
P_t(k) = \left ( b_m(k) + f_m(k)\mu^2 \right )^2P_m(k) \, ,
\label{eq:rsd}
\end{eqnarray}
where $f$ is the growth rate, not to be confused with the growth factor that is included in the transfer functions. The growth rate $f$ is defined as:
\begin{eqnarray}
f_m(k) = \frac{d \ln \left ( \sqrt{P_m(k,z)} \right )}{d\ln a} \, ,
\label{eq:fm}
\end{eqnarray}
with $a$ being the scale factor (of course, the growth rate is implicitly redshift-dependent). Finally, the $\mu$ term in Eq.~(\ref{eq:rsd}) is defined as the cosine of the angle between the Fourier mode $\mathbf{k}$ and the line-of-sight vector $\mathbf{r}$:
\begin{eqnarray}
\mu \equiv \frac{\mathbf{k} \cdot \mathbf{r}}{kr} \, , \quad k \equiv \vert \mathbf{k} \vert \, , \quad r \equiv \vert \mathbf{r} \vert \, .
\end{eqnarray}
Note that the expression in Eq.~(\ref{eq:rsd}) is not exact, as it lacks an exponential suppression due to the fingers-of-God (FoG) effect~\cite{Samushia:2011cs,Bull:2014rha}. If we make the simplifying assumption that the halo velocity dispersion $\sigma$ (which enters the FoG correction) does not depend on halo mass (and hence indirectly on the bias), the FoG correction then becomes bias-independent and therefore does not impact our discussion. Hence, for simplicity, we drop it for the moment. It will be included later in the analysis.

Kaiser's result was derived for models in which the growth rate is scale-independent, which is true for the minimal $\Lambda$CDM model, but not in presence of massive neutrinos. In the latter case, one could expect again that RSD effects are driven solely by the baryon and cold dark matter fluctuations. This has been checked explicitly on the basis of simulations in~\cite{Villaescusa-Navarro:2017mfx}, where it was proven that in the presence of RSD and massive neutrinos (but neglecting FoG effects), one has:
\begin{eqnarray}
P_t(k,M_{\nu}) = \left (b_{cb}(k) + f_{cb}(k,M_{\nu})\mu^2 \right )^2P_{cb}(k,M_{\nu}) \, , \nonumber \\
\label{eq:rsd_cb}
\end{eqnarray}
with:
\begin{eqnarray}
f_{cb}(k,M_{\nu}) \equiv \frac{d \ln \left ( \sqrt{P_{cb}(k,z,M_{\nu})} \right )}{d\ln a} \, .
\label{eq:fcb}
\end{eqnarray}
As in the previous subsection, we can always use this result to define an effective bias $b_m(k,M_{\nu})$ and an effective growth rate $f_m^{\rm eff}(k,z,M_{\nu})$ such that the following holds:
\begin{eqnarray}
P_t(k,M_{\nu}) = \left ( b_m(k,M_{\nu}) + f_m^\mathrm{eff} (k,z,M_{\nu})\mu^2 \right )^2P_m(k,M_{\nu}) \, , \nonumber \\
\label{eq:rsd2}
\end{eqnarray}
provided that the ``effective'' bias $b_m(k,M_{\nu})$ is related to the ``meaningful'' one by Eq.~(\ref{eq:bm}), while the effective growth factor should be defined as:
\begin{eqnarray}
f_m^{\rm eff}(k,z,M_{\nu}) \equiv \left ( \frac{T_{cb}(k,z,M_\nu)}{T_m(k,z,M_\nu)} \right ) f_{cb}(k,z,M_{\nu}) \, .
\label{eq:fmeff}
\end{eqnarray}
As a note of warning, the above should not be confused with the actual growth rate of the total matter spectrum:
\begin{eqnarray}
f_m(k,z,M_{\nu}) = \frac{d \ln \left ( \sqrt{P_m(k,z,M_{\nu})} \right )}{d\ln a} \, .
\label{eq:fm}
\end{eqnarray}
Both of these effective quantities, $b_m$ and $f_m^{\rm eff}$, are easy to implement at the level of likelihood modules of an MCMC sampling package, with a proper computation of $f_{cb}(k,z,M_{\nu})$ and $f_m^{\rm eff}(k,z,M_{\nu})$ based on the total matter spectrum and on transfer functions. However, as we shall discuss very shortly, it is better to perform an implementation directly at the level of the Einstein-Boltzmann solver, which can be modified to output directly 
$P_{cb}$ and $f_{cb}$ on top of $P_m$ and $f_{m}$: this allows to better take non-linear growth effects into account.

We emphasize that in the absence of RSD, the NISDB effect could be corrected equivalently by either rescaling the bias [Eq.~(\ref{eq:bm})] or the power spectrum [Eq.~(\ref{eq:pt_cb2})] by related quantities, $(T_{cb}/T_m)$ or $(T_{cb}/T_m)^2$ respectively. The symmetry between bias and power spectrum is broken by the growth rate term in the RSD correction.

\subsection{The impact of non-linearities}
\label{subsec:non-linear}

Before discussing the details and results of our analysis, the impact of non-linear evolution on cosmological perturbations needs to be discussed. Usually, in Boltzmann solvers, non-linear effects are introduced via prescriptions such as \texttt{Halofit}~\cite{Takahashi:2012em} or \texttt{HMcode}~\cite{Mead:2015yca}: these are accurate simulations-based fitting formulas for the non-linear matter power spectrum. In~\cite{Bird:2011rb} \texttt{Halofit} has been revised and extended to describe non-linear evolution in cosmologies with massive neutrinos.

Let us consider the case in which the NISDB correction, envisaged in Sec.~\ref{subsec:rsd}, is implemented at the level of likelihood module in the MCMC sampling package. In such a way, the correction would be implemented \textit{after} applying the \texttt{Halofit} prescription to the linear evolution. However, the transfer functions entering Eqs.~(\ref{eq:bm}, \ref{eq:fmeff}) are by definition linear quantities. Hence, the NISDB correction as described in Sec.~\ref{subsec:rsd}, would rescale the non-linear power spectrum by a linear quantity, and would be somewhat inconsistent.

One way to address this issue is to derive the right quantities not at the level of likelihood module, but rather at the level of the Einstein-Boltzmann solver. We modified the code \texttt{CLASS}~\cite{Blas:2011rf} in such a way to output at the same time $P_{cb}(k, z, M_\nu)$ and $P_{m}(k, z, M_\nu)$ for each given model. The former is used for computing the galaxy spectrum $P_t(k, z, M_\nu)$ [derived from Eqs.~(\ref{eq:rsd_cb}, \ref{eq:fcb})], while the latter can be used for computing other observables, for instance the Limber-approximated cosmic shear spectrum. For consistency, when computing non-linear spectra, the code processes the linear $P_{cb}(k, z, M_\nu)$ with a version of \texttt{Halofit} without massive neutrino corrections, and the linear $P_{m}(k, z, M_\nu)$ with a version of \texttt{Halofit} including the massive neutrino corrections~\footnote{Note that another consistent way of getting the non-linear total matter power spectrum $P_{m}(k, z, M_\nu)$, previously investigated in Ref.~\cite{Castorina:2015bma}, is to compute first the non-linear $P_{cb}(k)$ using \texttt{Halofit} without neutrino mass effects, and then to add the contribution of the linear neutrino spectrum $P_\nu(k)$ and of the cross-correlation term.} of Ref.~\cite{Bird:2011rb}. These new features with be part of a forthcoming release of the public \texttt{CLASS} code.

\subsection{Prescription Summary}
\label{subsec:recap}

Let us briefly summarize our prescription for correcting for the NISDB effect, taking RSD and non-linear effects correctly into account. The prescription works according to the following three steps:
\begin{enumerate}
\item Compute the CDM+baryons power spectrum $P_{cb}(k)$ by tracking the evolution of the CDM and baryon overdensities in the Einstein-Boltzmann solver.
\item Use the \texttt{Halofit} prescription to calculate the non-linear corrections to $P_{cb}(k)$, and obtain the non-linear CDM+baryons power spectrum $P_{cb}^{\rm nl}(k)$. Pay attention to which version of \texttt{Halofit} is used. The version used in this case should \textit{not} have the non-linear corrections due to massive neutrinos.
\item Finally, in the likelihood module, multiply the obtained non-linear CDM+baryons power spectrum $P_{cb}^{\rm nl}(k)$ by the \textit{universal} $b_{cb}$-dependent RSD correction, to obtain the theoretical galaxy power spectrum $P_t^{\rm th}(k)$, i.e. to first approximation:
\begin{eqnarray}
P_t^{\rm th}(k,M_{\nu}) = \left ( b_{cb}(k) + f_{cb}(k,M_{\nu})\mu^2 \right )^2 \, P_{cb}^{\rm nl}(k,M_{\nu}) \, . \nonumber \\
\label{pre}
\end{eqnarray}
The obtained theoretical galaxy power spectrum $P_t^{\rm th}(k)$ can then be compared to the measured galaxy power spectrum through the likelihood function, in order to obtain constraints on the cosmological parameters through the usual MCMC analysis. In principle, a non-linear RSD correction should be used instead of the linear Kaiser formula  in Eq.~(\ref{pre}): for instance, see~\cite{Matsubara:2007wj,Sato:2011qr} for a resummed perturbation theory approach to modeling non-linear RSD. However, the correct way of implementing non-linear RSD corrections is still a matter of debate in the community. In this work, we limit our analysis to linear scales and hence only model linear RSD (see Sec.~\ref{sec:datasets}).
\end{enumerate}

\section{Datasets and Analysis methodology}
\label{sec:datasets}

In this section, we discuss the analysis we conduct in order to determine the impact of the NISDB correction on constraints from future galaxy clustering data. For each experiment, we construct a mock likelihood wherein the role of the mock data spectrum is played by the theoretical spectrum of a fiducial cosmological model. These likelihoods take into account the expected noise level and systematic uncertainties associated with the experiments.

We perform an MCMC parameter inference from these combined likelihoods in order to forecast the sensitivity of the future experiments to the model parameters (this approach was already followed e.g. in~\cite{Perotto:2006rj,Audren:2012vy,Sprenger:2018tdb}). Our analysis pipeline is implemented in \texttt{MontePython}~\cite{Audren:2012wb}, an MCMC sampling and likelihood package that has recently been updated to \texttt{v3.0}~\cite{Brinckmann:2018cvx}. The features of \texttt{v3.0}, most notably an expanded suite of likelihoods and a more efficient Metropolis-Hastings sampler, were employed for this work. \texttt{MontePython} is interfaced with the Boltzmann solver \texttt{CLASS}~\cite{Blas:2011rf}. We choose a fiducial model where $M_{\nu}$ is pessimistically set to $0.06\,{\rm eV}$ (since this value of $M_{\nu}$ should in principle be the hardest to detect), the minimal value allowed by neutrino oscillation data in the normal ordering scenario~\cite{Gonzalez-Garcia:2015qrr,Capozzi:2016rtj,Esteban:2016qun,Capozzi:2017ipn,deSalas:2017kay}. 

On the CMB side, we use the likelihood \texttt{fake\_planck\_realistic}~\cite{DiValentino:2016foa} included in \texttt{MontePython} \texttt{v3.0}, taking into account temperature, polarisation and CMB lensing extraction. We adopt noise spectra roughly matching those expected from the full \textit{Planck} results~\footnote{Courtesy of Anthony Challinor, for the CORE-M5 proposal.}. For the purpose of forecasting sensitivities, it is easier to use a mock Planck likelihood rather than a real one, because we can then use the exact same fiducial model across all likelihoods.

On the side of Large Scale Structure (LSS), we employ the Fourier-space galaxy clustering mock likelihood \texttt{euclid\_pk} presented in~\cite{Sprenger:2018tdb}, with specifications for the \textit{Euclid} mission taken from~\cite{Amendola:2012ys,Geach:2009tm,Audren:2012vy}. The likelihood implements the following corrections and approximations (for full details see Sec.~2.1 of~\cite{Sprenger:2018tdb}):
\begin{itemize}
\item Redshift-space distortions using the Kaiser formula~\cite{Kaiser:1987qv}.
\item Exponential suppression due to Fingers of God~\cite{Bull:2014rha}.
\item Correction due to limited instrumental resolution.
\item Correction for Alcock-Paczy\'{n}ski effect.
\item Two nuisance parameters, $\beta_0$ and $\beta_1$, to account for inaccuracies in the bias evolution with redshift $b(z)=\beta_0 (1+z)^{0.5\beta_1}$.
\item Flat sky approximation~\cite{Lemos:2017arq,Asgari:2016txw}.
\item NISDB correction discussed in this work.
\end{itemize}

We remark that the procedure to correct for the NISDB outlined in Sec.~\ref{subsec:recap} is fully general once RSD and non-linear effects are properly accounted for. However we will only use the linear power spectrum in this work. Our choice is motivated by considerations of simplicity, in addition to the fact that the correct way of accounting for non-linear RSD is still under debate. To restrict ourselves to linear scales, we impose a cut-off $k_{\rm max}$ in wavenumber space~\footnote{We note that the \texttt{euclid\_pk}  likelihood features an option of imposing a theoretical error increasing with wavenumber $k$. This theoretical error mimics the effect of theoretical uncertainties in the modeling of non-linear effects anticipated for the analysis of Euclid data. In the present analysis, we did not use this tool in order to get more easily reproducible results.}, scaling with redshift as~\cite{Smith:2002dz}:
\begin{eqnarray}
k_{\rm max}(z) = k_{\rm max}(z=0) \times (1+z)^{\frac{2}{2+n_s}} \, ,
\end{eqnarray}
where $k_{\rm max}(z=0)=0.2\,h\,{\rm Mpc}^{-1}$ and $n_s$ is the tilt of the primordial scalar power spectrum. We note that it will be interesting to check the impact of including non-linear scales, modulo a correct modeling thereof, on our results, and defer this analysis to future work.

We consider a seven-parameter cosmological model described by the six-parameter concordance $\Lambda$CDM model with the addition of the sum of the neutrino masses $M_\nu$. The seven parameters are the physical baryon and CDM energy densities $\omega_b \equiv \Omega_b h^2$ and $\omega_{cdm} \equiv \Omega_{cdm} h^2$, the angular size of the sound horizon at decoupling $\theta_s$, the redshift of reionization $z_{\rm reio}$ in place of the optical depth to reionization $\tau_{\rm reio}$, and the amplitude and tilt of the primordial scalar power spectrum $A_s$ and $n_s$, in addition to the sum of the neutrino masses $M_{\nu}$. The neutrino mass spectrum is approximated as three mass eigenstates degenerate in mass. It has been shown that this approximation is sufficiently accurate for the sensitivity of current and future data, which are strongly sensitive to the sum of the neutrino masses $M_{\nu}$ and hardly sensitive to the masses of the individual eigenstates~\cite{Lesgourgues:2004ps,Lesgourgues:2005yv,DeBernardis:2009di,Jimenez:2010ev,Carbone:2010ik,Hall:2012kg,
Giusarma:2016phn,Gerbino:2016sgw,Archidiacono:2016lnv,DiValentino:2016foa,Vagnozzi:2017ovm}. The values of the cosmological parameters defining our fiducial model are reported in Tab.~\ref{tab:fiducials}.
\begin{table}[!h]
\centering
\begin{tabular}{|c||c|}
\hline
Parameter	& Fiducial value\\
\hline
$\omega_b$	&$0.02218$\\
$\omega_{cdm}$	&$0.1205$\\
$100\,\theta_s$	&$1.04146$\\
$z_{\rm reio}$	&$8.24$\\
$\ln[10^{10}A_s]$	&$3.056$\\
$n_s$	&$0.9619$\\
$M_{\nu} \, ({\rm eV})$	&$0.06$\\
\hline
\end{tabular}
\caption{Fiducial values for the cosmological parameters adopted when simulating future cosmological data.}
\label{tab:fiducials}
\end{table}

To explore the impact of not correctly accounting for the NISDB correction, we consider the following two cases, denoted by \textit{correct} and \textit{wrong} respectively:
\begin{itemize}
\item \textit{Correct}: the NISDB correction is properly applied, at the level of Boltzmann solver, as discussed in Sec.~\ref{subsec:recap}.
\item \textit{Wrong}: the NISDB correction is not applied at all. We use the same fiducial mock data as in the \textit{correct} case.
\end{itemize}
We run MCMC chains for the two cases considered above, monitoring the convergence of the generated chains through the Gelman-Rubin $R-1$ parameter~\cite{Gelman:1992zz}, and requiring $R-1<0.01$ for the chains to be considered converged.

\section{Results}
\label{sec:results}

\begin{figure}[!h]
\centering
\includegraphics[width=1.0\linewidth]{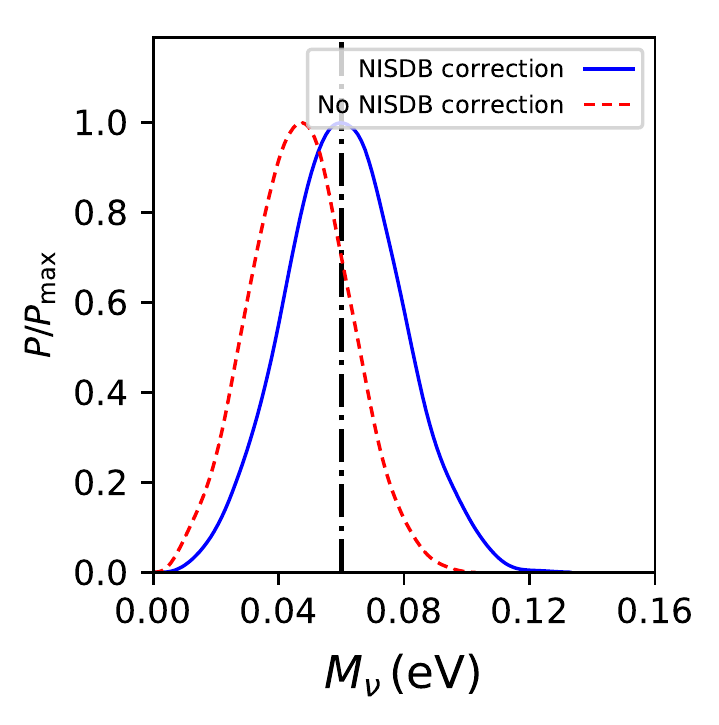}
\caption{One-dimensional marginalized posterior probabilities, normalized to their maximum values, of the sum of the active neutrino masses $M_{\nu}$ (in eV) for the two cases considered in this analysis: the \textit{correct} case (blue solid), where the neutrino-induced scale-dependent bias (NISDB) correction is properly applied, and the \textit{wrong} case (red dashed), where the NISDB correction is not applied. The dot-dashed vertical line at $M_{\nu}=0.06\,{\rm eV}$ shows the input fiducial value of $M_{\nu}$ used in our analysis. It is clearly visible that the input fiducial value is perfectly recovered for the blue curve (\textit{correct} case). When the correction is not applied, we fail in recovering the input fiducial value, as shown by the red curve.}
\label{fig:mnu}
\end{figure}

Here, we discuss the results of our MCMC analysis. The posterior distributions of $M_{\nu}$ are shown in Fig.~\ref{fig:mnu}, with the \textit{correct} case corresponding to the blue solid curve and the \textit{wrong} case to the red dashed curve. The vertical dot-dashed line at $M_{\nu}=0.06\,{\rm eV}$ corresponds to the input fiducial value of $M_{\nu}$. The 68\%~C.L. bounds on $M_\nu$ from the two cases are reported in Tab.~\ref{tab:bounds}. 

We first consider the \textit{correct} case, where the NISDB correction is properly applied. In this case, we can see that we correctly recover the fiducial value of $M_{\nu}$, represented by the vertical dot-dashed line in Fig.~\ref{fig:mnu}. In fact, we find $M_{\nu} = (0.061 \pm 0.019)\,{\rm eV}$, in perfect agreement with the fiducial value of $M_{\nu} = 0.06\,{\rm eV}$.

\begin{table}[!h]
\centering
\begin{tabular}{|c||c|c|}
\hline
Parameter	& NISDB correction & No NISDB correction\\
\hline
$M_{\nu} \, ({\rm eV})$	&$0.061 \pm 0.019$	&$0.046 \pm 0.015$\\
\hline
$\omega_{cdm} \equiv \Omega_{cdm}h^2$	&$0.1205 \pm 0.0003$	&$0.1207 \pm 0.0003$\\
$n_s$	&$0.9621 \pm 0.0014$	&$0.9612 \pm 0.0014$\\
\hline
\end{tabular}
\caption{Contraints at 68\%~C.L. on the sum of the neutrino masses $M_\nu$ and on the two cosmological parameters most correlated with $M_\nu$, the cold dark matter physical density $\omega_{cdm} \equiv \Omega_{cdm}h^2$ and the scalar spectral index $n_s$. The constraints are obtained from the combination of mock CMB and galaxy clustering data described in Sec.~\ref{sec:datasets}. The datasets are simulated to match the expected sensitivity of the final data release from the \textit{Planck} satellite and the upcoming \textit{Euclid} satellite. The constraints are reported for the two cases considered in this analysis: the correction for the neutrino-induced scale-dependent bias is applied; the correction is not applied. By comparing the limits with the input fiducial values in Tab.~\ref{tab:fiducials}, it clear that failure to apply the correction leads to biased determinations of cosmological parameters.}
\label{tab:bounds}
\end{table}

Next, we consider the \textit{wrong} case where the NISDB correction is not applied. In this case, we see that the choice of not applying the NISDB correction has biased our determination of $M_{\nu}$. We find $M_{\nu} = (0.046 \pm 0.015)\,{\rm eV}$, about $1\sigma$ away from the fiducial value of $M_{\nu} = 0.06\,{\rm eV}$. It is interesting to note that when the NISDB correction is not applied, the result is not only a biased determination of $M_{\nu}$, but it also features a spurious $\sim 25\%$ decrease in the error bar $\sigma_{M_{\nu}}$, consistent with the previous findings of~\cite{Raccanelli:2017kht}.

These results are consistent with analytical expectations. It is well-known that on linear scales the effect of non-zero neutrino masses is to suppress the total matter power spectrum by an amount approximately given by~\cite{Hu:1997mj,Lesgourgues:2006nd,Lesgourgues:1519137}:
\begin{eqnarray}
\frac{P_m(k,f_{\nu})}{P_m(k,f_{\nu}=0)} \simeq 1-8f_{\nu} \, ,
\label{eq:suppression1}
\end{eqnarray}
where $f_{\nu}$ is the neutrino fraction:
\begin{eqnarray}
f_{\nu} \equiv \frac{\rho_{\nu}}{\rho_{\nu}+\rho_c+\rho_b} = \frac{\Omega_{\nu}}{\Omega_m} \, .
\end{eqnarray}
However, the CDM+baryons power spectrum is actually reduced by a smaller amount:
\begin{eqnarray}
\frac{P_{cb}(k,f_{\nu})}{P_{cb}(k,f_{\nu}=0)} \simeq 1-6f_{\nu} \, .
\label{eq:suppression2}
\end{eqnarray}
Comparing Eq.~(\ref{eq:suppression1}) and Eq.~(\ref{eq:suppression2}), we see that the impact of not applying the NISDB correction is to first approximation expected to lead to a decrease in both the inferred mean value $M_{\nu}$ and error $\sigma_{M_{\nu}}$ by a factor of $8/6$. This ratio approximately matches our results for $M_{\nu}^{{\rm correct}}/M_{\nu}^{\rm wrong}=0.061\,{\rm eV}/0.046\,{\rm eV}$ and $\sigma_{M_{\nu}}^{{\rm correct}}/\sigma_{M_{\nu}}^{{\rm wrong}}=0.019\,{\rm eV}/0.015\,{\rm eV}$.

\begin{figure}[!htb]
\centering
\includegraphics[width=1.0\linewidth]{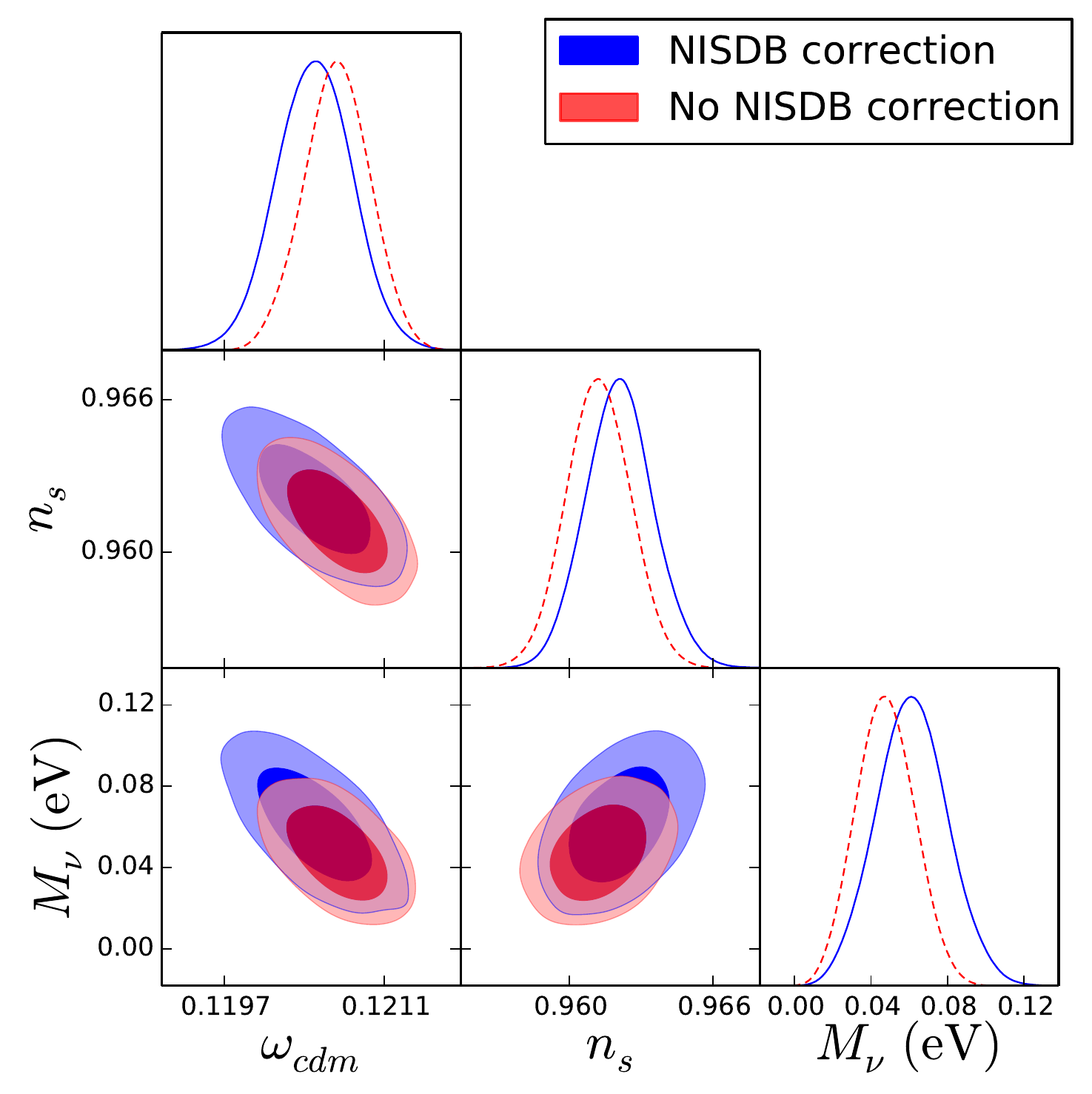}
\caption{Triangular plot showing joint and one-dimensional marginalized posterior distributions for the parameters which are most correlated with the sum of the three active neutrino masses $M_{\nu}$ (in eV). These parameters are the dark matter physical density $\omega_{cdm} \equiv \Omega_{cdm} h^2$ and the scalar spectral index $n_s$. The panels along the diagonal show the one-dimensional probability distributions of the individual parameters. The remaining blocks show the 2D joint distributions. The distributions are shown for two cases: the case where the neutrino-induced scale-dependent bias (NISDB) correction is properly applied (blue, 1D posteriors are given by solid curves), and the case where the NISDB correction is not applied (red, 1D posteriors are given by dashed curves). The one-dimensional distributions along the diagonal represent normalized probability distributions and are hence in arbitrary units.}
\label{fig:tri}
\end{figure}

Finally, shifts in the inferred value of $M_{\nu}$ are expected to impact the inferred values of other cosmological parameters which are degenerate with $M_{\nu}$. In particular, we have checked that the two most affected parameters are the CDM physical energy density $\omega_{cdm} \equiv \Omega_{cdm}h^2$ and the scalar spectral index $n_s$. The CDM physical energy density is negatively correlated with $M_{\nu}$, while $n_s$ is positively correlated. Both degeneracies are well-understood and documented in the literature~\cite{Gerbino:2016sgw,Archidiacono:2016lnv}. In Fig.~\ref{fig:tri} we show a triangular plot featuring the joint and one-dimensional posterior distributions of $M_{\nu}$, $\omega_{cdm}$, and $n_s$, for both the case where the NISDB correction is applied (blue solid curves/blue contours) and the case where it is not applied (red dashed curves/red contours).

The 68\%~C.L. bounds on $\omega_{cdm}$ and $n_s$ are reported in Tab.~\ref{tab:bounds}. For the \textit{correct} case, where the NISDB correction is applied, we find $\omega_{cdm} = 0.1205 \pm 0.0003$, perfectly recovering the input fiducial value. Similarly, we find $n_s=0.9621 \pm 0.0014$, also perfectly recovering the input fiducial value. For the \textit{wrong} case where the NISDB correction is not applied, the two previous values shift to $\omega_{cdm} = 0.1207 \pm 0.0003$ and $n_s = 0.9612 \pm 0.0014$ respectively. Although these correspond to $<1\sigma$ shifts, they provide further indications that implementing the NISDB correction is important, not only for future determinations of the total neutrino mass, but also of other cosmological parameters, as found in~\cite{Raccanelli:2017kht}.

Similar considerations concerning shifts in other parameters would hold in extended cosmologies as well, especially when considering additional parameters which are to some extent degenerate with $M_{\nu}$ (such as the dark energy equation of state parameter $w$ and the curvature density parameter $\Omega_k$), and could be explored in future work. As a final remark, we remind the reader that the details concerning the shifts and direction of degeneracy between the discussed parameters depend to some extent on the type of data used. For instance, when baryon acoustic oscillation distance measurements are considered, the degeneracy between $M_{\nu}$ and $n_s$ is expected to be reverted~\cite{Gerbino:2016sgw}, i.e. the two parameters become negatively correlated instead of positively correlated.

\section{Conclusions}
\label{sec:conclusions}

Cosmological data is exquisitely sensitive to the sum of the three active neutrino masses $M_{\nu}$, and a combination of measurements from next-generation surveys is expected to provide the first measurement ever of $M_{\nu}$, and thus of the absolute neutrino mass scale. Galaxy clustering is particularly sensitive to the effects of non-zero $M_{\nu}$. However, galaxy clustering analyses also present significant challenges, such as the correct modeling of galaxy bias. Failure to do so could introduce significant model systematics which propagate to the determination of cosmological parameters, including $M_{\nu}$. It is known that massive neutrinos introduce a scale-dependence in the galaxy bias \textit{even on large scales}, if the bias is defined with respect to the total matter field. On the other hand, the bias defined with respect to the cold dark matter plus baryons field is \textit{universal}, hence independent of the effects of $M_{\nu}$. Most cosmological analyses in the presence of massive neutrinos so far have ignored this effect, defining the bias with respect to the total matter field and, at the same time, treating the bias as universal. Not accounting for this \textit{neutrino-induced scale-dependent bias} (NISDB) could introduce severe model systematics in future analyses of galaxy clustering data.

In this work, we have quantified the importance of properly correcting for the NISDB effect when analysing galaxy clustering data. This issue was previously addressed in~\cite{Raccanelli:2017kht} using a Fisher matrix forecast. We revisit it through an MCMC sensitivity forecast and with an extended modeling of systematic effects. We have presented a simple prescription for correcting for the NISDB effect, summarized in Sec.~\ref{subsec:recap}. In doing so, we have also clarified some subtle issues concerning the correct way to implement the NISDB correction in the presence of redshift-space distortions and non-linearities.

We then presented a forecast based on mock cosmic microwave background and large-scale structure likelihoods, intended to mimic the legacy data release from the \textit{Planck} satellite and measurements of the galaxy power spectrum from the \textit{Euclid} satellite. We have shown that failure to implement the NISDB correction can introduce systematics in the inferred value of  $M_{\nu}$. In particular, we find that the value of $M_{\nu}$ inferred is a factor of $\sim 8/6$ lower than the fiducial value. At the same time, the $1\sigma$ uncertainty on the inferred value decreases by the same factor of $\sim 8/6$ with respect to the case where the correction is properly implemented. The latter effect represents a spurious increase in sensitivity. These results agree with the findings of~\cite{Raccanelli:2017kht} and match theoretical expectations. Finally, we have examined how the shift in the inferred value of $M_{\nu}$ correspondingly propagates to shifts in other cosmological parameters, such as $\Omega_{cdm} h^2$ and $n_s$.

We encourage the community to correctly account for the NISDB effect in future analyses of galaxy clustering data in the presence of massive neutrinos, in order to increase the robustness of the analyses and minimize the impact of modeling systematics. The tools necessary to easily correct for the neutrino-induced scale-dependent bias effect will be made publicly available in an upcoming release of the \texttt{CLASS} code. \\

\textbf{\textit{Note added.}} Our choice of title is inspired by a previous work on the issue of the neutrino-induced scale-dependent bias by Raccanelli et al.~\cite{Raccanelli:2017kht}, Shakespearianly titled \textit{``Bias from neutrino bias: to worry or not to worry?''}, with an obvious quotation from Hamlet's soliloquy. As our paper is somewhat complementary to~\cite{Raccanelli:2017kht}, we have also chosen to go for a title inspired by a quote from the same play, namely \textit{``Madness in great ones must not unwatch'd go''}, a quote by King Claudius from \textit{Hamlet}.

\begin{acknowledgments}
S.V. thanks Juli\'{a}n B. Mu\~{n}oz for enlightening correspondence and discussions. T.B. is supported by the Deutsche Forschungsgemeinschaft through the graduate school ``Particle and Astroparticle Physics in the Light of the LHC'' and through the individual grant ``Cosmological probes of dark matter properties''. K.F., M.G., and S.V. acknowledge support by the Vetenskapsr\aa det (Swedish Research Council) through contract No. 638-2013-8993 and the Oskar Klein Centre for Cosmoparticle Physics. K.F. acknowledges support from DoE grant de-sc0007859 at the University of Michigan as well as support from the Leinweber Center for Theoretical Physics. This work has been done thanks to the facilities offered by the RWTH High Performance Computing cluster under project \texttt{rwth0113}, and the National Energy Research Scientific Computing Center (NERSC).
\end{acknowledgments}

\vskip 0.3 cm

\bibliographystyle{JHEP}
\bibliography{bias.bib}

\end{document}